\pdfoutput=1

\documentclass[%
 amsmath,amssymb,
 aps,
 prl, 
 a4paper, 
 twocolumn, 
 showkeys
]{revtex4-1}

\usepackage{graphicx}
\usepackage{dcolumn}
\usepackage{bm}

\renewcommand{\vec}[1]{{\bf #1}}
\newcommand{\braket}[1]{\langle #1  \rangle} 

\begin{document}


\title{A Variational Approach to Monte Carlo Renormalization Group}

\author{Yantao Wu$^1$ and Roberto Car$^{1, 2}$}

\affiliation{%
$^1$The Department of Physics, Princeton University\\
$^2$The Department of Chemistry, Princeton University \\
}%

\date{\today}

\begin{abstract}
We present a Monte Carlo method for computing the renormalized coupling constants and the critical exponents within renormalization theory. 
The scheme, which derives from a variational principle, overcomes critical slowing down, by means of a bias potential that renders the coarse grained variables uncorrelated. The 2D Ising model is used to illustrate the method. 
\end{abstract}

\pacs{Valid PACS appear here}
\maketitle


Since the introduction of renormalization group (RG) theory \cite{wilson_rg}, there has been strong interest in methods to compute the renormalized coupling constants and the critical exponents in a non-perturbative fashion. 
This goal has been achieved with the Monte Carlo (MC) RG approach of Swendsen. 
In 1979, he introduced a method to compute the critical exponents, which did not require explicit knowledge of the renormalized Hamiltonian \cite{mcrg}. 
A few years later, he solved the problem of calculating the renormalized coupling constants, using an equality due to Callen \cite{callen} to write the correlation functions in a form explicitly depending on the couplings. 
By imposing that the standard MC expression of a correlation function and its corresponding Callen form be equal, he derived equations whose iterative solution led to the coupling constants \cite{mcrg_rc}. 
Finding the renormalized Hamiltonian is an example of inverse statistical mechanical problem \cite{invising}. 
MCRG has been used successfully in many applications but difficulties related to sampling efficiency may be severe. 
Typically, the evaluation of the correlation functions near a critical point suffers from critical slowing down and is affected by large sampling errors in large systems. 
This difficulty can be alleviated with ingenious cluster algorithms \cite{clustermc}, which, however, are limited to specific models.   

Here we present an MCRG framework based on a variational principle for a biasing potential acting on the coarse grained degrees of freedom of a RG transformation. 
In our approach, the coupling constants and the critical exponents derive from the same unifying principle. 
Swendsen's formulae emerge as a special case, but our scheme also leads to formulations exempt from critical slowing down. 
In addition, it permits to estimate variationally the effect of truncating the Hamiltonian.  

Although the approach is rather general, here we limit ourselves, for concreteness, to lattice models with discrete spin degrees of freedom, $\{\bm \sigma\}$. 
A generic Hamiltonian has the form  
\begin{equation}
\label{eq:hamiltonian} 
H(\bm \sigma) = \sum_{\alpha} K_\alpha S_\alpha(\bm \sigma), 
\end{equation}
where the $K_\alpha$ are coupling constants and the $S_\alpha$ are operators acting on the spins $\bm\sigma$, such as sums or products of spins or combinations thereof.  

RG considers a flux in the space of Hamiltonians (\ref{eq:hamiltonian}) under scale transformations that reduce the linear size of the original lattice by a factor $b$. 
The rescaled degrees of freedom take the same discrete values of the original spins, to which they are related by a coarse graining transformation, $\bm\sigma' = \tau(\bm\sigma)$. 
For example, $\tau$ can be the block spin transformation of Kadanoff \cite{block2}.   

The distribution of the $\bm\sigma'$ is obtained from the distribution of the $\bm\sigma$ by tracing out the original degrees of freedom while keeping the $\bm \sigma'$ fixed:  
\begin{equation}
  \label{eq:distribution}
  p(\bm \sigma') = \frac{\sum_{\bm \sigma}  \delta_{\bm \tau(\bm\sigma), \bm\sigma'} e^{-H(\bm \sigma)}}{Z} =  \frac{e^{-H'(\bm \sigma')}}{Z'} . 
\end{equation}
Here $\delta$ is the discrete Kroneker-delta function, 
$Z$ and $Z'$ are partition functions that ensure the normalization of the corresponding distributions. 
While the partition function $Z'$ is invariant under RG transformations, the renormalized Hamiltonian $H'$ is not, except at fixed points of the RG flow:    
\begin{equation}
  Z = \sum_{\bm \sigma} e^{-H(\bm \sigma)} = \sum_{\bm\sigma'} e^{-H(\bm \sigma')} = Z'
\end{equation}
and 
\begin{equation}
  H'(\bm\sigma') = -\log \sum_{\bm\sigma} \delta_{\tau(\bm \sigma), \bm\sigma'} e^{-H(\bm\sigma)}
\end{equation}
Repeated {\it at infinitum}, the RG transformations generate a flux in the space of Hamiltonians, in which all possible coupling terms appear, unless forbidden by symmetry. 
For example, in an Ising model with no magnetic field, only even spin products appear. 
The space of the coupling terms is, in general, infinite. 
However, perturbative and non-perturbative calculations suggest that only a finite number of couplings should be sufficient for a given degree of accuracy. 

In the proximity of a critical point, the distribution (\ref{eq:distribution}) of the block spins $\bm\sigma'$ displays a divergent correlation length, originating critical slowing down of local MC updates. 
This can be avoided by modifying the distribution of the $\bm\sigma'$ by adding to the Hamiltonian $H'(\bm\sigma')$ a biasing potential $V(\bm\sigma')$ to force the biased distribution of the block spins, $p_V(\bm\sigma')$, to be equal to a chosen {\it target distribution}, $p_t(\bm\sigma')$. For instance, $p_t$ can be the constant probability distribution.  
Then the $\bm\sigma'$ have the same probability at each lattice site and act as uncorrelated spins, even in the vicinity of a critical point.           

It turns out that $V(\bm\sigma')$ obeys a powerful variational principle that facilitates the sampling of the Landau free energy \cite{varyfes}. 
In the present context, we define the functional $\Omega[V]$ of the biasing potential $V(\bm\sigma')$ by: 
\begin{equation}
\Omega [V] = \log\frac{ \sum_{\bm \sigma'} e^{-[H'(\bm \sigma') + V(\bm \sigma')]}}{\sum_{\bm \sigma'} e^{-H'(\bm \sigma')}}+ \sum_{\bm \sigma'} p_t (\bm \sigma') V(\bm \sigma'), 
\end{equation}
where $p_t(\bm \sigma')$ is a normalized known target probability distribution. As demonstrated in \cite{varyfes}, the following properties hold:  
\begin{enumerate}
\item $\Omega [V]$ is a convex functional with a lower bound. 
\item The minimizer, $V_{\text{min}}(\bm\sigma')$, of $\Omega$ is unique up to a constant and is such that: 
\begin{equation}
\label{eq:fes}
H'(\bm \sigma') = - V_{\text{min}}(\bm \sigma') - \log p_t (\bm \sigma') + \text{constant} 
\end{equation}
\item The probability distribution of the $\bm\sigma'$ under the action of $V_{\text{min}}$ is:   
\begin{equation}
  p_{V_{\text{min}}}(\bm \sigma') = \frac{e^{-(H'(\bm \sigma') + V_{\text{min}}(\bm \sigma'))}}{\bm \sum_{\sigma'} e^{-(H'(\bm\sigma') + V_{\text{min}} (\bm\sigma'))}} =  p_t(\bm \sigma')
\end{equation}
\end{enumerate}
The above three properties lead to the following MCRG scheme.  

First, we approximate $V(\bm\sigma')$ with $V_{\vec J}(\bm\sigma')$, a linear combination of a finite number of terms $S_\alpha(\bm\sigma')$ with unknown coefficients $J_\alpha$, forming a vector $\vec J = \{J_1, ..., J_\alpha, ..., J_n\}$.    
\begin{equation}
  V_{\vec J}(\bm\sigma') = \sum_\alpha J_\alpha S_\alpha(\bm\sigma') 
\end{equation}
Then the functional $\Omega[V]$ becomes a convex function of $\vec J$, due to the linearity of the expansion, and the minimizing vector, $\vec J_{\text{min}}$, and the corresponding $V_{\text{min}}(\bm\sigma')$ can be found with a local minimization algorithm using the gradient and the Hessian of $\Omega$: 
\begin{equation}
\label{eq:gradient}
\frac{\partial \Omega(\vec J)}{\partial J_\alpha} = - \braket{S_\alpha(\bm \sigma')}_{V_{\vec J}} + \braket{S_\alpha(\bm \sigma')}_{p_t}
\end{equation}
\begin{equation}
\label{eq:hessian}
\frac{\partial^2 \Omega(\vec J)}{\partial J_\alpha \partial J_\beta} = \braket{S_\alpha(\bm \sigma') S_\beta(\bm \sigma')}_{V_{\vec J}} - \braket{S_\alpha(\bm \sigma')}_{V_{\vec J}}\braket{S_\beta(\bm \sigma')}_{V_{\vec J}}
\end{equation}
Here $\braket{\cdot}_{V_{\vec J}}$ is the biased ensemble average under $V_{\vec J}$ and $\braket{\cdot}_{p_t}$ is the ensemble average under the target probability distribution $p_t$. 
The first average is associated to the Boltzmann factor $\exp\{-(H'(\bm \sigma') + V(\bm \sigma'))\} = \sum_{\bm\sigma} \delta_{\tau(\bm\sigma), \bm\sigma'} \exp(-H(\bm\sigma)) \exp(-V(\tau(\bm\sigma)))$ and can be computed with MC sampling. 
The second average can be computed analytically if $p_t$ is simple enough.      
$\braket{\cdot}_{V_{\vec J}}$ always has inherent random noise, or even inaccuracy, and some sophistication is required in the optimization problem. 
Following \cite{varyfes}, we adopt the stochastic optimization procedure of \cite{bach}, and improve the statistics by running independent MC simulations, called {\it multiple walkers}, in parallel. 
For further details, consult \cite{varyfes} and the  Supplementary Material (SM) \cite{sm}. 

The renormalized Hamiltonian $H'(\bm\sigma')$ is given by Eq. \ref{eq:fes} in terms of $V_{\text{min}}(\bm\sigma')$. Taking a constant $p_t$, we have modulo a constant: 
\begin{equation}
  H'(\bm \sigma') = -V_{\text{min}}(\bm\sigma') = \sum_{\alpha} (-J_{\text{min}, \alpha}) S_\alpha(\bm \sigma') 
\end{equation}
In this finite approximation the renormalized Hamiltonian has exactly the same terms of $V_{\text{min}}(\bm \sigma')$ with renormalized coupling constants 
\begin{equation}
  K'_\alpha = -J_{\text{min}, \alpha}. 
\end{equation}
The relative importance of an operator $S_\alpha$ in the renormalized Hamiltonian can be estimated variationally in terms of the relative magnitude of the coefficient $J_{\text{min}, \alpha}$. 
When $J_{\text{min}, \alpha}$ is much smaller than the other components of $\vec J_{\text{min}}$, the corresponding $S_\alpha(\bm\sigma')$ is comparably unimportant and can be ignored. The accuracy of this approximation could be quantified by measuring the deviation of $p_{V_{\text{min}}}(\bm\sigma')$ from $p_t(\bm\sigma')$.  

To illustrate the method, we present a study of the Ising model on a $2D$ square lattice in the absence of a magnetic field. 
We adopt $3 \times 3$ block spins with the majority rule. 
26 coupling terms were chosen initially, including 13 two-spin and 13 four-spin products. 
One preliminary iteration of variational RG (VRG) was performed on a $45\times 45$ lattice starting from the nearest-neighbor Hamiltonian.
The coupling terms with renormalized coupling constants smaller than 0.001 in absolute value were deemed unimportant and dropped from further calculations. 
13 coupling terms, including 7 two-spin and 6 four-spin products, survived this criterion and were kept in all subsequent calculations \cite{sm}. 
Each calculation consisted of 5 VRG iterations starting with nearest-neighbor coupling, $K_{nn}$, only. 
All the subsequent iterations used the same lattice of the initial iteration. 
Standard Metropolis MC sampling \cite{metropolis} was adopted, and the calculations were done at least twice to ensure that statistical noise did not alter the results significantly. 

\begin{figure}[hth]
\centering
\includegraphics[scale=1]{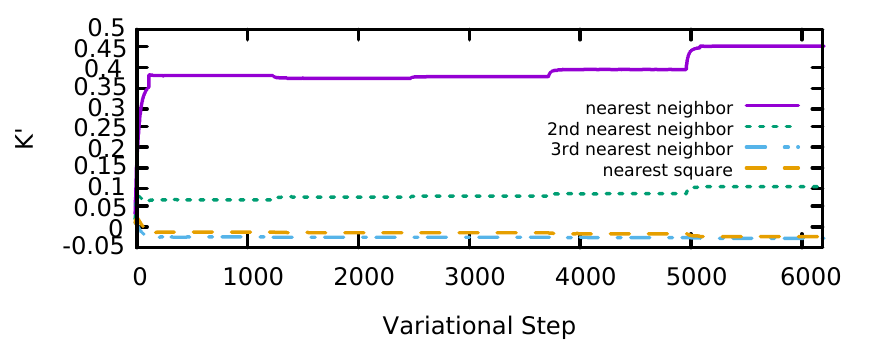}
\includegraphics[scale=1]{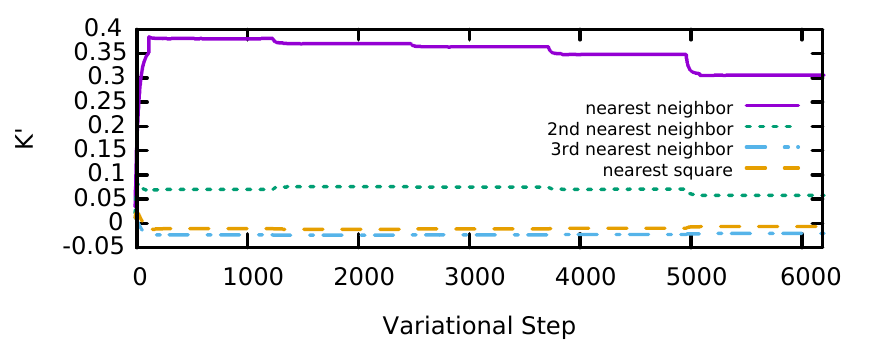}
\caption{(color online). Variation of the renormalized coupling constants over five VRG iterations on a $300\times300$ lattice. 
Each iteration has 1240 variational steps, each consisting of 20 MC sweeps. 
16 multiple walkers are used for the ensemble averages in Eqs. \ref{eq:gradient} and \ref{eq:hessian}. 
For clarity, we only show the four largest renormalized couplings after the first iteration. 
Full plots are reported in the SM \cite{sm}.  
Top: Simulation starting with $K_{nn} = 0.4365$. 
Bottom: Simulation starting with $K_{nn} = 0.4355$.}
\label{fig:300_rg}
\end{figure}

In Fig. \ref{fig:300_rg}, results are shown for a $300 \times 300$ lattice with two initial $K_{nn}$, equal to $0.4355$ and to $0.4365$, respectively. When $K_{nn} = 0.4365$, the renormalized coupling constants increase over the five iterations shown, and would increase more dramatically with further iterations. 
Similarly, they decrease when $K_{nn} = 0.4355$. 
Thus, the critical coupling $K_c$ should belong to the window $0.4355-0.4365$. 
The same critical window is found for the $45\times45$, $90\times 90$, $150\times 150$, and $210\times 210$ lattices \cite{sm}.  
Because each iteration is affected by truncation and finite size errors, less iterations for the same rescaling factor would reduce the error.
For example, 4 VRG iterations with a $2\times2$ block have the rescaling factor of a $16 \times 16$ block. 
The latter is computationally more costly than a calculation with $2\times 2$ blocks, but can still be performed with modest computational resources. Indeed, with a $16 \times 16$ block, RG iterations on a $128 \times 128$ lattice gave a critical window $0.4394-0.4398$ \cite{sm}, very close to the exact value, $K_c \sim0.4407$, due to Onsager \cite{onsager}.

The statistical uncertainty of the renormalized couplings from the variational method is small. 
Using the standard approach, Ref. \cite{mcrg2dising} found a renormalized nearest neighbor coupling equal to $0.408 \pm 0.002$ after the first RG iteration on a $36\times 36$ lattice using a $3 \times 3$ block spin, starting with $K_{nn} = 0.4407$. 
This result required $5.76 \times 10^5$ MC sweeps. 
With our method, applied to a $300 \times 300$ lattice, starting with $K_{nn} = 0.4365$, we found a renormalized nearest-neighbor coupling equal to $0.38031 \pm 0.00002$ after $3.398 \times 10^5$ MC sweeps.
The standard error in our case was computed with the block averaging method \cite{block_method}. Because \cite{mcrg2dising} used only seven coupling terms and a different initial $K_{nn}$, the renormalized couplings should not be expected to be the same in the two calculations, but a comparison of the corresponding statistical uncertainties should be meaningful.  

According to theory \cite{wilsonkondo}, the critical exponents are obtained from the leading eigenvalues of $\frac{\partial K'_\alpha}{\partial K_\beta}$, the Jacobian matrix of the RG transformation, at a critical fixed point. 
In order to find $\frac{\partial K'_\alpha}{\partial K_\beta}$ near a fixed point, we need to know how the renormalized coupling constants $K'_\alpha$ from a RG iteration on the Hamiltonian $H = \sum_\beta K_\beta S_\beta$, change when $K_\beta$ is perturbed to $K_\beta + \delta K_\beta$, for fixed target probability $p_t$ and operators $S_\alpha$. 
The minimum condition, Eq. \ref{eq:gradient}, 
implies $\frac{d\Omega}{d J_\alpha} = 0$, i.e. for all $\gamma$:  
\begin{equation}
\frac{\sum_{\bm \sigma} S_\gamma(\bm\sigma') e^{- \sum_\beta (K_\beta S_\beta(\bm\sigma) - K'_\beta S_\beta(\bm\sigma'))}}{\sum_{\bm\sigma} e^{- \sum_\beta (K_\beta S_\beta(\bm\sigma) - K'_\beta S_\beta(\bm\sigma'))}}  = \braket{S_\gamma(\bm \sigma')}_{p_t}, 
\end{equation}
and  
\begin{equation}
\label{eq:2ndcondition}
\begin{split}
\frac{\sum_{\bm \sigma} S_\gamma(\bm\sigma') e^{- \sum_\beta ((K_\beta + \delta K_\beta) S_\beta(\bm\sigma) - (K'_\beta + \delta K'_\beta) S_\beta(\bm\sigma'))}}{\sum_{\bm\sigma} e^{- \sum_\beta ((K_\beta + \delta K_\beta) S_\beta(\bm\sigma) - (K'_\beta + \delta K'_\beta)S_\beta(\bm\sigma'))}}  \\
= \braket{S_\gamma(\bm \sigma')}_{p_t}.
\end{split} 
\end{equation}
Expanding Eq. \ref{eq:2ndcondition} to linear order in $\delta K'_\alpha$ and $\delta K_\beta$, we obtain (\cite{sm}) 
\begin{equation}
\label{eq:matrix_eq}
A_{\beta\gamma} = \sum_\alpha \frac{\partial K'_\alpha}{\partial K_\beta} \cdot  B_{\alpha\gamma}, 
\end{equation}
where 
\begin{equation}
\label{eq:A}
A_{\beta\gamma} = \braket{S_\beta(\bm \sigma) S_\gamma(\bm \sigma')}_{V} - \braket{S_\beta(\bm \sigma)}_{V}\braket{S_\gamma(\bm\sigma')}_{V}, 
\end{equation}
and 
\begin{equation}
\label{eq:B}
B_{\alpha\gamma} = \braket{S_\alpha(\bm \sigma') S_\gamma(\bm \sigma')}_{V} - \braket{S_\alpha(\bm\sigma')}_{V}\braket{S_\gamma(\bm \sigma')}_{V}. 
\end{equation}
Here $\braket{\cdot}_V$ denotes average under the biased Hamiltonian, $\widetilde{H} = \sum_\beta K_\beta S_\beta(\bm\sigma) - K'_\beta S_\beta(\bm\sigma')$.

If we require the target average of $S_\gamma(\bm\sigma')$ to coincide with the unbiased average under $H = \sum_\beta K_\beta S_\beta$, $K'$ would necessarily vanish and Eqs. \ref{eq:A}-\ref{eq:B} would coincide with Swendsen's formulae \cite{mcrg}. 
If we use a uniform target probability, the $\bm \sigma'$ at different sites would be uncorrelated, and critical slowing down would be absent.  

In practice, in order to compute the critical exponents, we first need to locate $K_c$. 
From the above calculations on the $45 \times 45$, $90 \times 90$, and $300 \times 300$ lattices with a $3 \times 3$ block spin, we expect that $K_c = 0.436$ should approximate the critical nearest-neighbor coupling in our model. 
Indeed an RG iteration starting from this value gives couplings that remain essentially constant, as illustrated in Figs. S11-S13 of the SM \cite{sm}. 

\begin{figure}[hth]
\centering
\includegraphics[scale=1]{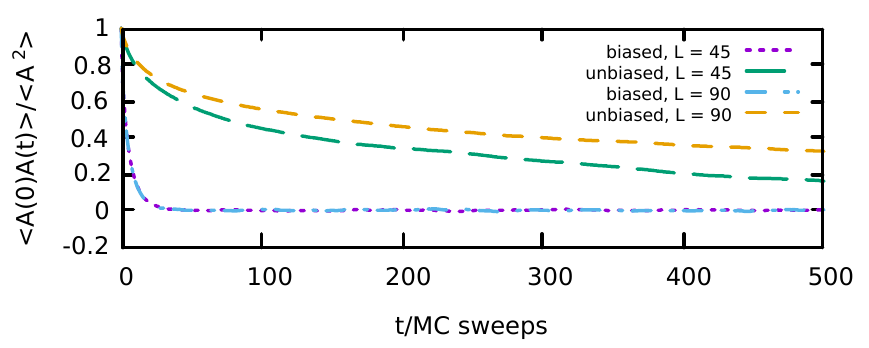}
\caption{(color online). Time correlation of the estimator $A = S_0(\bm\sigma)S_0(\bm\sigma')$ on $45\times45$ and $90\times 90$ lattices (Eq. \ref{eq:A}). $S_0$ is the nearest neighbor term in the simulations of Table \ref{table:critical}.}
\label{fig:cortime}
\end{figure}

Then, we use Eqs. \ref{eq:matrix_eq}-\ref{eq:B} to compute the Jacobian of the RG transformation by setting $K_c = 0.436$. 
The renormalized coupling constants after the first RG iteration represent $K_\alpha$, and those after the second RG iteration represent $K'_\alpha$. 
The results for biased and unbiased ensembles are shown in Table \ref{table:critical}, which reports the leading even ($e$) and odd ($o$) eigenvalues of $\frac{\partial K'_\alpha}{\partial K_\beta}$ when including 13 coupling terms for the three $L\times L$ lattices with $L = 45, 90$, and $300$.  
As seen from the table, biased and unbiased calculations give slightly different eigenvalues, as one should expect, given that the respective calculations are different embodiments of the truncated Hamiltonian approximation. 
For $L = 300$ the results are well converged in the biased ensemble.  
By contrast, we were not able to obtain converged results for this lattice in the unbiased ensemble on the time scale of our simulation.    
The absence of critical slowing down in the biased simulation is demonstrated in Fig. \ref{fig:cortime}, which displays  time decay of a correlation function in the biased and unbiased ensembles. See also Figs. S14-S15 of the SM \cite{sm}. 
\begin{table}[htb]
  \setlength{\tabcolsep}{1em}
  \begin{tabular}{l l l l} 
    \hline
    \hline
      &$L$ & $\lambda_1^e$ & $\lambda_1^o$\\
    \hline
    unbiased & 45 & $2.970(1)$ & 7.7171(2) \\
    &90 & $2.980(3)$ &  7.7351(1)\\

    biased & 45 & $3.045(5)$ & 7.858(4)  \\
    &90 & $3.040(7)$ &  7.870(2)\\
    &300 & $3.03(1)$ &  7.885(5)\\

    Exact &  & 3  & 7.8452 \\
    \hline
    \hline
  \end{tabular}
  \caption{Leading even (e) and odd (o) eigenvalues of $\frac{\partial K'_\alpha}{\partial K_\beta}$ at the approximate fixed point found with VRG, in both the unbiased and biased ensembles. 
The number in parentheses is the statistical uncertainty on the last digit, obtained from the standard error of 16 independent runs.  
13 (5) coupling terms are used for even (odd) interactions.   
The calculations used $10^6$ MC sweeps for the $45 \times 45$ and $90\times 90$ lattices, and $5 \times 10^5$ sweeps for the $300 \times 300$ lattice.
}
  \label{table:critical}
\end{table}

The fixed point used for Table \ref{table:critical} is approximate, and we did not make any effort to fine tune the approximation. 
Refinements could be done iteratively using Eqs. \ref{eq:matrix_eq}-\ref{eq:B}, as we will discuss in a future paper. 
There is an important benefit in knowing accurately the location of the fixed point, because then a single RG iteration, instead of multiple implicit iterations would suffice to compute the Jacobian.  
Moreover, one could use small block spins, having a smaller statistical uncertainty than larger block spins. 

In summary, we have unified the calculation of critical exponents and renormalized couplings within the same framework.
A key feature of our approach is that we adopt a biased ensemble, $\braket{\cdot}_V$, for the averages. 
This not only simplifies the algorithm, but also enhances the sampling. 
In fact, the original motivation for the variational principle \cite{varyfes} was to overcome the long correlation time in first-order phase transitions. 
The bias potential constructed by optimizing the functional acquires a history-dependence that discourages the sampling of previously visited configurations \cite{varyfes}, thereby breaking the long correlation time of the unbiased simulation. 
In the RG context, enhanced sampling eliminates critical slowing down.  
We expect that it should be also helpful in systems with deep local free energy minima, as the variational method was originally designed to deal precisely with such systems. 

The finite size of the numerical samples is a source of error.
If the RG iterations are carried out on a single $L\times L$ lattice, the coarse grained lattice will have size $\frac{L}{b} \times \frac{L}{b}$. 
Then, as noted in \cite{mcrg2dising}, the calculated renormalized couplings will have different size errors on the $L \times L$ and $\frac{L}{b} \times \frac{L}{b}$ lattices. 
A better way, as suggested in \cite{two_lattice}, would be to perform calculations on two lattices, $L \times L$ and $\frac{L}{b} \times \frac{L}{b}$, so that the coarse grained lattice rescaled by $b^n$, at the $n$th iteration starting from $L \times L$, would coincide with the lattice rescaled by $b^{n-1}$, at the $(n-1)$th iteration starting from $\frac{L}{b} \times \frac{L}{b}$. 
In this way, two successive RG iterations have the same lattice size, with a significant cancellation of finite size errors. 
We plan to discuss in a future paper how this idea could be implemented within VRG.  

In the present paper we have used a constant probability distribution $p_t$, but there is no reason to always do so. 
For example, in systems with continuous and unbounded degrees of freedom, like molecular systems or lattice field theory, it may be convenient to use a Gaussian distribution for $p_t$. 

Finally, we note that a regular term $g(K)$ always appears as the inhomogeneous part of a RG transformation \cite{nauenberg}:  
\begin{equation}
  \exp{[H'(K'; \bm \sigma') + Ng(K)]} = \sum_{\bm\sigma} \delta_{\tau(\bm\sigma), \bm\sigma'}\exp{[H(K; \bm\sigma)]}
\end{equation}
The $g(K)$ in this equation is precisely the thermodynamic free energy per site in the biased ensemble $\braket{\cdot}_V$, as shown in the SM \cite{sm}.
It is then interesting, and somewhat surprising, that the information on the critical behavior is fully contained in the statistical behavior of $\braket{\cdot}_V$, even though $g(K)$ is a regular function and $\braket{\cdot}_V$ does not show singular behavior.           

All the codes used in this project were written in C\texttt{++}, and would be available upon request. The authors would like to thank C. Castellani and L. Pietronero for discussions. Partial support for this work was provided by the Department of Energy under Grant no. DE-FG02-05ER46201. 
\bibliographystyle{apsrev}

\end{document}